\documentclass[useAMS,usenatbib]{mn2e}
\usepackage{bm}
\usepackage{macros}
\usepackage[usenames,dvipsnames,svgnames,table]{xcolor}
\usepackage{hyperref}
\definecolor{darkblue}{rgb}{0.0,0.0,0.3}
\hypersetup{colorlinks,breaklinks,
            linkcolor=darkblue,urlcolor=darkblue,
            anchorcolor=darkblue,citecolor=darkblue}
\usepackage{graphicx}
\usepackage{amsmath}
\usepackage{epstopdf}

\def\ba{\begin{eqnarray}}
\def\ea{\end{eqnarray}}
\def\be{\begin{equation}}
\def\ee{\end{equation}}
\def\vph{\varphi}

\title[On the primary beam deceleration in the pulsar wind]{On the primary beam 
deceleration in the pulsar wind}
\author[L. I. Arzamasskiy, V. S. Beskin, V. V. Prokofev]{L. I. Arzamasskiy$^{1}$\thanks{E-mail:
lev.arzamasskiy@phystech.edu, beskin@lpi.ru}, 
V. S. Beskin$^{1,2}$\footnotemark[1], V. V. Prokofev$^1$\\
$^{1}$Moscow Institute of Physics and Technology, Dolgoprudny, Institutsky per., 9, 
Moscow region, 141700, Russia\\
$^{2}$P.N.Lebedev Physical Institute, Leninsky prosp., 53, Moscow, 119991, Russia}

\begin{document}

\date{Accepted. Received; in original form}

\pagerange{\pageref{firstpage}--\pageref{lastpage}} \pubyear{2015}

\maketitle

\label{firstpage}

\begin{abstract}
We investigate the motion of the primary beam outside the light cylinder in the pulsar 
wind. Inside the light cylinder both primary and secondary 
plasma move along dipole magnetic field lines where their energies can be arbitrary. 
But at larger distances the theory predicts quasi-radial motion with the velocity 
exactly corresponding to the drift velocity which cannot be the same for primary 
and secondary plasma. Hence, the deceleration of the primary beam is to take place 
simultaneously resulting in the acceleration of the secondary plasma. We investigate this 
process in the three-fluid MHD approximation and demonstrate that for most pulsars the 
energy of the beam remains practically unchanged. Only for young radio pulsars (Crab, 
Vela) essential deceleration up to the energy of the secondary plasma takes place outside the 
fast magnetosonic surface $r_{\rm F} \sim (10$--$100) R_{\rm L}$, the energy of secondary 
plasma itself increasing insufficiently. 
\end{abstract}


\section{Introduction}
\label{sect:intro}

According to the common point of view, the activity of radio pulsars is connected 
with the electron-positron plasma generated near magnetic poles~\citep{1977puls.book.....M, 
1977puls.book.....S}. Numerous works devoted to this subject~\citep{sturrock71,rs75, 
1981ApJ...248.1099A, daugherty1982pulsarcascades, 1985ZhETF..89....3G, 2007AstL...33..660I, 
2007MNRAS.382.1833M, 2013MNRAS.429...20T} demonstrate that the outflowing plasma consists 
of the primary beam with the Lorentz-factor
$\gamma^{\rm b} \sim 10^{7}$ and the number density $n^{\rm b} \approx n_{\rm GJ}$,
where $n_{\rm GJ} = \Omega B/2 \pi c e$ is the Goldreich-Julian number density, and 
the secondary electron-positron plasma with $n^{\pm} \sim (10^3$--$10^5) n_{\rm GJ}$,
and $\gamma^{\pm} \sim 10^2$. 

It is clear that inside the light cylinder $r < R_{\rm L}= c/\Omega$ both the primary
beam and the secondary plasma move along the dipole magnetic field lines where their 
energies can be arbitrary. On the other hand, at large enough distances from a neutron 
star $r \gg R_{\rm L}$  both the analytical theory~\citep{michel_pulsar_wind_1994, bes98} 
and numerical simulations ~\citep{ckf99, 2006ApJ...648L..51S, SashaMHD, 2014ApJ...785L..33P} 
predict the quasi-radial outflow of the relativistic electron-positron plasma along the 
poloidal magnetic field with the radial velocity exactly corresponding to the drift motion 
\begin{equation}
{\bf U}_{\rm dr} = c \frac{{\bf E} \times {\bf B}}{B^2}.
\label{drift}
\end{equation}
Clearly, this drift velocity cannot describe simultaneously the motion to the beam and the
secondary plasma which have sufficiently different energies. 

Thus, the question arises about the energy evolution of these two components as they escape 
the pulsar magnetosphere and propagate outwards forming the pulsar wind. We  show that 
efficient deceleration of primary beam can be realized for young energetic radio pulsars 
(Crab, Vela) only. In this case the deceleration up to the energy of the secondary plasma 
takes place outside the fast magnetosonic surface \mbox{$r_{\rm F} \sim (10$--$100) R_{\rm L}$,} 
the energy of secondary plasma itself increases insufficiently. For ordinary pulsars with period 
$P \sim 1$~s, the energy of the primary beam (due to escaping to the region with rather small 
electric and magnetic fields) remains practically unchanged. 

The paper is organized as follows. We start with the description of basic equations 
of three-fluid MHD in Sect.~\ref{sect:BE}. In Sect.~\ref{sect:GP} we describe the 
general properties of these equations. The beam damping is discussed in 
Sect.~\ref{sect:BD}. Finally, in Sect.~\ref{sect:res} the main results of our
consideration are formulated.

\section{Basic Equations}
\label{sect:BE}

In this section we derive the basic three-fluid MHD equations. Throughout the paper 
we use the spherical coordinate system $(r,\theta,\vph)$ with unit vectors 
$({\bf e}_r,{\bf e}_\theta,{\bf e}_\vph)$. All quantities having the superscript 
'$\rm b$' correspond to the primary beam, and the quantities with superscript '$\pm$' 
refer to the secondary plasma particles (electrons or positrons). Finally, electron 
charge equals $-e$.

To describe the deceleration of the primary beam (and possible acceleration of the secondary
plasma) we start from the well-known \citet{michel_pulsar_wind_1994} monopole force-free solution
\begin{align}
B_r & =  B_{\rm L}\frac{R_{\rm L}^2}{r^2},  
\label{leqf1} \\
E_{\theta} & =  B_{\varphi} = -B_{\rm L}\frac{R_{\rm L}}{r} \sin\theta,
\label{leqf2} 
\end{align}
corresponding to the zero particle mass $m_{\rm e}$. Within this approximation, the particles
move radially with the velocity $v = c$, so that the current density $j_{r} = c \rho_{\rm e}$ and
the charge density
\begin{equation}
\rho_{\rm e} = -\frac{\Omega B_r}{2 \pi c} \cos\theta
\label{rhoe}
\end{equation}
just corresponding to Goldreich-Julian charge density
\begin{equation}
\rho_{\rm GJ} = -\frac{\bf \Omega B}{2 \pi c}. 
\end{equation}

One can easily check that the fields (\ref{leqf1})--(\ref{leqf2}) with the charge density
$\rho_{\rm e}$ (\ref{rhoe}) and the current density $j_{r} = c \rho_{\rm e}$ are the 
exact solutions of the time-independent Maxwell equations
\begin{align}
\label{Max1}
\nabla\cdot{\bf E}& = 4\pi\rho_{\rm e},\hspace{8pt}\nabla\times{\bf E} = 0,
\\ 
\label{Max2}
\nabla\cdot{\bf B} &= 0,\hspace{22pt}
\nabla\times{\bf B} = \frac{4\pi}{c}{\bf j}.
\end{align}
For this reason one can model the outflow with the massless particles moving radially with the
velocity ${\bf v} = c \, {\bf e}_{r}$ and having the following number density
\begin{align}
n^{\pm} & =  \lambda \frac{\Omega B_{\rm L}}{2\pi c e}\frac{R_{\rm L}^2}{r^2},
\label{feq}\\
n^{\rm b} & =  \frac{\Omega B_{\rm L}}{2\pi c e}\frac{R_{\rm L}^2}{r^2}\cos \theta .
 \end{align}
Here $\lambda = en_{\rm e}/|\rho_{\rm GJ}| \gg 1$ is the so-called multiplication parameter 
which equals $10^3$--$10^5$ for most radio pulsars. In what follows we consider for simplicity the
case of $\lambda =$~const. Such a choice corresponds to 
the equality of the secondary electron and positron number densities. 

As one can see, the only difference with our previous  works~\citep{2000MNRAS.313..433B, 
beszak04} is that we consider here not two-component (i.e., secondary electron-positron 
plasma) but three-component outflow containing also the primary beam having Goldreich-Julian 
number density $n^{\rm b} = |\rho_{\rm GJ}|/e$ and Lorentz-factor
\mbox{$\gamma^{\rm b} \sim 10^7 \gg \gamma^{\pm}$.} As is shown in Appendix~\ref{LC}, 
this value is in agreement with the radiation reaction force acting on the beam particle 
as it moves along dipole magnetic field line inside the light cylinder. On the other hand, 
as is shown in Appendix~\ref{BI}, two-stream instability is not effective, and, hence, 
three-fluid hydrodynamical approximation under consideration is good enough to describe
the evolution of the primary beam. 

Then, following the papers mentioned above we can include into consideration the finite 
particle mass as a small disturbance of the force-free solution. One can do it because for 
magnetically dominated outflow $\sigma \gg 1$, the outflow remains 
actually radial~\citep{1994PASJ...46..123T, bes98}. Here
\begin{equation}
\sigma = \frac{\Omega e B_{\rm L} R_{\rm L}^2}{4\lambda m c^3}
\end{equation}
is the Michel magnetization parameter. For ordinary pulsars 
($P \sim 1$~s, $B_0 \sim 10^{12}$~G), one has $\sigma \sim 10^{3} - 10^{4}$, 
and only for the fast ones ($P \sim 0.1 - 0.01$~s, $B_0 \sim 10^{13}$ G), 
$\sigma \sim 10^{5} - 10^{6}$.

As a result, to find the structure of the flow we have to analyze Maxwell equations 
(\ref{Max1})--(\ref{Max2}) and the equation of motion for all three components:
\begin{align}
({\bf v}^{\pm},\nabla){\bf p}^{\pm}
&= \pm e\left({\bf E}+\frac{{\bf v}^{\pm}}{c}\times{\bf B}\right),
\label{Mot1} \\
({\bf v}^{\rm b},\nabla){\bf p}^{\rm b} &= 
 -e\left({\bf E}+\frac{{\bf v}^{\rm b}}{c}\times{\bf B}\right).
\label{Mot2}
\end{align}
It is convenient to introduce electric potential $\Phi_{\rm e}(r,\theta)$ and the 
magnetic flux function $\Psi(r,\theta)$, so that
\begin{align}
\label{Epol}
{\bf E} &= -\nabla \Phi_{\rm e}(r,\theta),\\
{\bf B}_{\rm p} &= \frac{\nabla\Psi\times{\bf e}_{\varphi}}{2\pi r\sin\theta}.
\label{Bpol}
\end{align}
Then, one can seek the solution in the form
 \begin{eqnarray}
 n^{\pm}  &=&  \frac{\Omega B_{\rm L}}{2\pi c e}\frac{R_{\rm L}^2}{r^2}
\left[\lambda+\eta^{\pm}(r,\theta)\right],
\label{feq}\\
n^{\rm b}  &=&  \frac{\Omega B_{\rm L}}{2\pi c e}\frac{R_{\rm L}^2}{r^2}
\left[\cos\theta + \eta^{\rm b}(r,\theta)\right], \\
\Phi_{\rm e}(r,\theta)  &=&  \frac{\Omega R_{\rm L}^2
B_{\rm L}}{c}\left[-\cos\theta+\delta(r,\theta)\right],\\
\Psi(r,\theta)  &=& 2\pi B_{\rm L} R_{\rm L}^2\left[1-\cos\theta+\varepsilon
f(r,\theta)\right],\\
v^{\pm,{\rm b}}_r  &=&  c\left[1-\xi^{\pm,{\rm b}}_r(r,\theta)\right],
\label{defvr}\\
v^{\pm,{\rm b}}_{\theta,\vph}  &=&
c\xi^{\pm,{\rm b}}_{\theta,\vph}(r,\theta)\label{defvth},
\end{eqnarray}
with the small disturbances $\eta$, $\delta$, $\varepsilon f$ and $\xi$ corresponding 
to the small perturbations caused by the finite particle mass.
Accordingly, using (\ref{Epol}) and (\ref{Bpol}), one can write down
 \begin{eqnarray}
B_r & = & B_{\rm L}\frac{R_{\rm L}^2}{r^2}\left(
1+\frac{\varepsilon}{\sin\theta}
\frac{\partial f}{\partial \theta}\right),\\
B_{\theta} & = &  -\varepsilon B_{\rm L} \frac{R_{\rm L}^2}{r \sin\theta}
\frac{\partial f}{\partial r},\\
B_{\varphi} & = & B_{\rm L}\frac{\Omega R_{\rm L}}{c}\frac{R_{\rm L}}{r}\left[-\sin\theta
-\zeta(r,\theta)\right], \\
E_r & = &  -B_{\rm L}\frac{\Omega R_{\rm L}^2}{c}\frac{\partial \delta}{\partial r},\\
E_{\theta} & = & B_{\rm L}\frac{\Omega R_{\rm L}^2}{c r}\left(
-\sin\theta - \frac{\partial \delta}{\partial \theta}\right).
\label{leq} 
\end{eqnarray}
Here all deflecting functions are supposed to be $\ll 1$. For $v_r=c$, $v_{\theta}=0$, and
$v_{\vph}=0$ for all types of particle, we return to the force-free Michel solution. 

Now, substituting Eqns. (\ref{feq})--(\ref{leq}) into Maxwell equtions (\ref{Max1})--(\ref{Max2}), 
we obtain to the first-order approximation the following system of linear equations:
\begin{align}
\label{G}
&2(\Delta\eta +\eta^{\rm b}) + \frac{\partial}{\partial r}\left(r^2
\frac{\partial \delta}{\partial r}\right)+
\frac{1}{\sin\theta}\frac{\partial}{\partial \theta}
\left(\sin\theta \frac{\partial\delta}{\partial \theta}\right)=0,\\
\label{Fr}
&\frac{1}{2\sin\theta}\frac{\partial}{\partial\theta}
(\zeta\sin\theta)=
\lambda\Delta\xi_r-\xi^{\rm b}_r \cos\theta-\Delta\eta - \eta^{\rm b},\\
 \label{FTh}
&\frac{\partial\zeta}{\partial r}=\frac{2}{r}
\left(\lambda\Delta\xi_{\theta}-\xi_{\theta}^{\rm b}\cos\theta\right),\\
\label{FPh}
\begin{split}
&\frac{\varepsilon}{\sin\theta}\frac{\partial^2 f}{\partial r^2}
+\frac{\varepsilon}{r^2}\frac{\partial}{\partial\theta}
\left(\frac{1}{\sin\theta}\frac{\partial f}
{\partial \theta}\right) = 
 \\
&~~~~~~~~~~~~~~~~~~~~~~~~~~~~~~~=2\frac{\Omega}{r c}\left(\cos\theta\xi^{\rm b}_{\varphi}
-\lambda\Delta\xi_{\varphi}\right).
\end{split}
\end{align}
Accordingly. equations of motion (\ref{Mot1}) and (\ref{Mot2}) now looks like
\begin{align}
\begin{split}
\label{MoTh1}
&\frac{\partial}{\partial r}\left(\xi_{\theta}^{\pm}\gamma^{\pm}\right)+
\frac{\xi_{\theta}^{\pm}\gamma^{\pm}}{r}=
\\ 
&=\pm 4\lambda\sigma\left(
-\frac{1}{r}\frac{\partial\delta}{\partial\theta}\right.\left.+
\frac{\zeta}{r}-\frac{\sin\theta}{r}\xi_r^{\pm}+
\frac{c}{\Omega r^2}\xi_{\varphi}^{\pm}\right), 
\end{split}
\end{align}
 \begin{align}
 \begin{split}
&\frac{\partial}{\partial r}\left(\xi_{\theta}^{\rm b}\gamma^{\rm b}\right)+
\frac{\xi_{\theta}^{\rm b}\gamma^{\rm b}}{r}=
\label{MoTh2}  \\
&~~~~~~~=-4\lambda\sigma\left(
-\frac{1}{r}\frac{\partial\delta}{\partial\theta}\right.\left.+
\frac{\zeta}{r}-\frac{\sin\theta}{r}\xi_r^{\rm b}+
\frac{c}{\Omega r^2}\xi_{\varphi}^{\rm b}\right), 
\end{split}\\
&\frac{\partial}{\partial r}\left(\gamma^{\pm}\right)=
\pm 4\lambda\sigma\left(
-\frac{\partial\delta}{\partial r}-
\frac{\sin\theta}{r}\xi_{\theta}^{\pm}\right), 
\label{MoR1} \\
&\frac{\partial}{\partial r}\left(\gamma^{\rm b}\right)
= -4\lambda\sigma\left(
-\frac{\partial\delta}{\partial r}-
\frac{\sin\theta}{r}\xi_{\theta}^{\rm b}\right), 
\label{MoR2} \\
\begin{split}
&\frac{\partial}{\partial r}\left(\xi_{\varphi}^{\pm}\gamma^{\pm}\right)+
\frac{\xi_{\varphi}^{\pm}\gamma^{\pm}}{r} =\label{MoPh1}  \\
&~~~~~~~~~~~~~~~~~~~~=\mp 4\lambda\sigma\left(
\varepsilon\frac{c}{\Omega r \sin\theta}\frac{\partial f}
{\partial r}\right.\left.+\frac{c}{\Omega r^2}\xi_{\theta}^{\pm}\right), 
\end{split}\\
\begin{split}
&\frac{\partial}{\partial r}\left(\xi_{\varphi}^{\rm b}\gamma^{\rm b}\right)+
\frac{\xi_{\varphi}^{\rm b}\gamma^{\rm b}}{r}=
\\
&~~~~~~~~~~~~~~~~~~~~=4\lambda\sigma\left(
\varepsilon\frac{c}{\Omega r \sin\theta}\frac{\partial f}
{\partial r}+\frac{c}{\Omega r^2}\xi_{\theta}^{\rm b}\right).
 \label{MoPh2}
 \end{split}
\end{align}
Here $\Delta A=A^+-A^-$, so these terms vanish in one fluid MHD description.

Formally, the system of equations (\ref{G})--(\ref{MoPh2}) requires fifteen 
boundary conditions. We consider for simplicity the case $\Omega R/c \ll 1$ 
when the star radius $R$ is much smaller than the light cylinder. As a result, 
one can write down the first nine boundary conditions as
\begin{eqnarray}
\xi_{\theta}^{\pm}(R_{\rm L},\theta) & = & 0, \\
\xi_{\theta}^{\rm b}(R_{\rm L},\theta) & = & 0,\\
\xi_{\varphi}^{\rm b}(R_{\rm L},\theta) & = & 0, \\
\xi_{\varphi}^{\pm}(R_{\rm L},\theta) & = & 0, \\
\gamma^{\pm}(R_{\rm L}, \theta) & = & \gamma_{\rm in},\\
\gamma^{\rm b}(R_{\rm L}, \theta) & = & \gamma_{\rm in}^{\rm b},
\end{eqnarray}
where $\gamma_{\rm in}$ and $\gamma_{\rm in}^{\rm b}$ are the initial Lorentz-factors 
of secondary plasma and beam particles correspondingly. 

Finally, as a boundary conditions one can put
 \begin{eqnarray}
\delta(R_{\rm L},\theta)  & = & 0,
\label{bc1}\\
\varepsilon f(R_{\rm L},\theta) & = & 0,
\label{bc2}\\
\eta^+(R_{\rm L},\theta)-\eta^-(R_{\rm L},\theta) & = & 0, \\
\eta^{\rm b}(R_{\rm L},\theta) & = & 0,
\end{eqnarray}
resulting from the relation
${\bf E}_{\rm L}
+ (\Omega R_{\rm L}/c){\bf e}_{\varphi} \times {\bf B}_{\rm L} = 0$
which implies a rigid rotation and a perfect conductivity of the surface
of a star. It is necessary to stress that all the expressions are 
valid only for a quasi-monopole outflow which exists only outside the 
light cylinder.

\section{General properties}
\label{sect:GP}

It is clear that in general case the analytical solution of Eqns. (\ref{G})--(\ref{MoPh2}) 
cannot be found. On the other hand, this system of equation contains some integrals of motion
which allow us to obtain the necessary information about the energy of the outflowing particles.
In particular, as was demonstrate by \citet{2000MNRAS.313..433B}, in the two-fluid approximation
this approach reproduces well-known one-fluid MHD results for the position of the fast 
magnetosonic surface for $\gamma_{\rm in} \ll \sigma^{1/3}$
\be 
r_{\rm F}  = \sigma^{1/3} \sin^{-1/3}\theta \, R_{\rm L}
\label{rF}
\ee
and the Lorentz-factor at this surface $\gamma_{\rm F} = \gamma^{\pm}(r_{\rm F})$
\be 
\gamma_{\rm F} = \sigma^{1/3}\sin^{2/3}\theta.
\label{gF}
\ee
Below we show that it is true for $\gamma_{\rm in} \gg \sigma^{1/3}$ as well.

First of all, excluding $\xi^{\pm}_{\theta}$ and $\xi^{\rm b}_{\theta}$ from Eqn.
(\ref{FTh}) and using (\ref{MoR1}) and (\ref{MoR2}) we obtain the following equation
\begin{align}
\frac{\partial\zeta}{\partial r}&= \frac{2}{\tan\theta}\frac{\partial\delta}{\partial r} - \nonumber\\
&-\frac{1}{2\sigma\sin\theta}\left(\frac{\partial\gamma^+}{\partial r}
+\frac{\partial\gamma^-}{\partial r}\right)
-\frac{1}{2\sigma\lambda\tan\theta}\frac{\partial \gamma^{\rm b}}{\partial r}.
\end{align}
It gives
\begin{eqnarray}
\zeta-\frac{2}{\tan\theta}\delta
+\frac{\lambda(\gamma^+ +\gamma^-)
+(\gamma^{\rm b}-\gamma_{\rm in}^{\rm b})\cos\theta}{2\lambda\sigma\sin\theta} =
\nonumber\\
= \frac{1}{\sigma\sin\theta}\gamma_{\rm in}+\frac{l(\theta)}{\sin\theta},
\label{EnCon}
\end{eqnarray}
where $\zeta(0,\theta)=l(\theta)/\sin\theta$ and $l(\theta)$ describes the disturbance 
of the electric current. Expression (\ref{EnCon}) corresponds to the conservation of the 
total energy flux along a magnetic field line. On the other hand, combining Eqns. 
(\ref{MoR1})--(\ref{MoPh2}), one can obtain the expressions corresponding
to conversation of the $z$-component of the angular momentum for all types of particles
\begin{align}
\delta & =  \varepsilon f \mp \frac{1}{4\lambda\sigma}\gamma^\pm
\left(1-\frac{\Omega r\sin\theta}{c}\xi_{\varphi}^\pm\right)
\pm \frac{1}{4\lambda\sigma}\gamma_{\rm in},
\label{AMP}\\
\delta & =  \varepsilon f+\frac{1}{4\lambda\sigma}\gamma^{\rm b}
\left(1-\frac{\Omega r\sin\theta}{c}\xi_{\varphi}^{\rm b}\right)
-\frac{1}{4\lambda\sigma}\gamma_{\rm in}^{\rm b}.
\label{AMB}
\end{align}

Now, neglecting the difference between $\gamma^{+}$ and $\gamma^{-}$
so that $\xi^+ = \xi^-$ and $\gamma^+ = \gamma^- = \gamma$
(which can be done in case of $\sigma \gg 1$ and $\lambda \gg 1$),
we obtain
\begin{equation}
\delta  =  \varepsilon f,
\end{equation}
and, hence,
\begin{equation}
\gamma\left(1-\frac{\Omega r\sin\theta}{c}\xi_{\varphi}\right)=\gamma_{\rm in}.
\label{WGD}
\end{equation}
Accordingly, from (\ref{AMB}) we have
\be
\gamma^{\rm b}\left(1-\frac{\Omega r\sin\theta}{c}\xi^{\rm b}_{\varphi}\right)
=\gamma_{\rm in}^{\rm b}.
\label{BGD}
\ee
On the other hand, Eqn. (\ref{MoR2}) for $\lambda\sigma \gg 1$ gives
\begin{eqnarray}
-\frac{1}{r}\frac{\partial\delta}{\partial\theta}+
\frac{\zeta}{r}-\frac{\sin\theta}{r}\xi_r+
\frac{c}{\Omega r^2}\xi_{\varphi}=0.
\label{ThF} 
\end{eqnarray}
Further, using the definitions (\ref{defvr})--(\ref{defvth}) to describe $\gamma$ 
in terms of $\xi_r$ and $\xi_{\theta}$, and neglecting $\xi_{\theta}$, one can write down 
\be
\gamma^2=\frac{1}{2\xi_r-\xi_{\varphi}^2}.
\label{GD}
\ee
As a result, combining relations (\ref{WGD}), (\ref{ThF}), (\ref{GD}), and (\ref{GD}), 
one can finally find the following algebraic equation on the hydrodynamical Lorentz-factor
of the secondary plasma $\gamma$
\begin{equation}
2\gamma^3-2\sigma\left[K+\frac{1}{2 x^2}
+\frac{\gamma_{\rm in}}{\sigma}\right]\gamma^2
+\sigma\sin^2\theta+\sigma\frac{c^2\gamma_{\rm in}^2}{\Omega^2r^2}= 0,
\label{GE}
\end{equation}
where $x=r/R_{\rm L}$ and
\begin{equation}
K(r,\theta)=2\cos\theta\delta-\sin\theta\frac{\partial\delta}{\partial\theta}
-\frac{\gamma^{\rm b}-\gamma_{\rm in}^{\rm b}}{2\sigma\lambda}\cos\theta+\frac{l(\theta)}{\sin\theta}.
\label{KD}
\end{equation}

Equations (\ref{GE})--(\ref{KD}) generalize the appropriate ones obtained by 
\citet{2000MNRAS.313..433B} for two-component plasma and for $\gamma_{\rm in} \ll \sigma^{1/3}$. 
As was already stressed, these equations allow us to find the position of
fast magnetosonic surface corresponding to the intersection of two roots of algebraic 
equation (\ref{GE}). Indeed, determining the derivative $r\partial\gamma/\partial r$, 
one can obtain
\be
r\frac{\partial\gamma}{\partial r}
=\gamma \sigma \frac{r\partial K/\partial r-x^{-2}[1-(\gamma_{\rm in}/\gamma)^2]}
{3\gamma-2\gamma_{\rm in}-\sigma(2K+x^{-2})}.
\ee
As the fast magnetosonic surface is the $X$-point, both the numerator and denominator 
are both equal to zero there.  As a result, evaluating $r\partial K/\partial r$ as $K$,
we obtain the following well-known asymptotic solutions \citep{2001A&A...371.1155B}:
\begin{enumerate}
\item
Fast rotator ($\gamma_{\rm in}<\sigma^{1/3}$). In this case we may neglect the terms 
containing $\gamma _{\rm in}$. It gives for $\delta_{\rm F} = \delta(r_{\rm F})$, etc.
\begin{eqnarray}
\delta_{\rm F} & \approx & \sigma^{-2/3},\label{DelS} \\
r_{\rm F} & \approx & \sigma^{1/3}\sin^{-1/3}\theta\,R_{\rm L},
\label{RS} \\
\gamma_{\rm F} & =& \sigma^{1/3}\sin^{2/3}\theta,
\end{eqnarray}
the last expression being exact. As we see, for fast rotator the acceleration of the
secondary plasma from $\gamma = \gamma_{\rm in}$ to $\gamma = \sigma^{1/3}$ takes place
up to the fast magnetosonic surface.
\vspace{0.5cm}
\item
Slow rotator ($\gamma_{\rm in}>\sigma^{1/3}$). In this case we obtain
\begin{eqnarray}
\delta_{\rm F} & \approx &  \gamma_{\rm in}\sigma^{-1},
\label{DelG} \\
r_{\rm F} & \approx & \sigma^{1/2}\gamma_{\rm in}^{-1/2} \,R_{\rm L},
\label{RG}\\
\gamma_{\rm F} & \approx & \gamma_{\rm in}.
\end{eqnarray}
For slow rotator there is no particle acceleration within the fast magnetosonic surface
which has a spherical shape.
\end{enumerate}

\section{Beam Damping and Plasma Acceleration}
\label{sect:BD}

\subsection{Beam damping}

In this section we determine how the energy of the primary beam changes 
as it outflows from the pulsar magnetosphere. As we consider the secondary 
plasma in the MHD limit where there is no longitudinal electric field, we 
may neglect the term $\partial \delta/\partial r$. This leads to the simple 
determination of $\xi_\theta^{\rm b}$ from Eqn. (\ref{MoR2}):
\begin{equation}
\xi_\theta^{\rm b} \approx \frac{r}{4\lambda \sigma}\frac{\partial 
\gamma^{\rm b}}{\partial r},
\end{equation}
where we also consider the case $\sin\theta \sim \cos \theta \sim 1$. On
the other hand, for \mbox{$\lambda \sigma \gg 1$} one can set the quantity 
in brackets in Eqn. (\ref{MoTh1}) equal to zero. After that we can rewrite 
Eqn. (\ref{MoTh2}) in the following way:
\begin{equation}
\frac{\partial}{\partial r} (r \xi^{\rm b}_\theta \gamma^{\rm b}) 
= 4 \lambda \sigma \left[\xi_r^{\rm b}-\xi_r - 
\frac{r}{R_{\rm L}}(\xi_\vph^{\rm b}-\xi_\vph)\right].
\end{equation}

Now, using Eqs. (\ref{WGD})--(\ref{BGD}), (\ref{GD}) and the definition
\begin{equation}
\xi_r^{\rm b} = \frac{1}{2 (\gamma^{\rm b})^2} + \frac{(\xi_\vph^{\rm b})^2}{2},
\end{equation}
one can finally obtain the following general equation describing the beam damping:
\begin{eqnarray}
\label{eqn:damp}
\frac{\partial}{\partial x} \left(x^2 \frac{\partial p}{\partial x}\right)
 = \Lambda ^2\left(\frac{1}{(\gamma^{\rm b}_{\rm in})^2}\frac{1}{p+1} 
- \frac{1}{\gamma^2}-\frac{1}{x^2} \frac{p}{p+1}+\frac{\Delta}{x^2}\right).
\end{eqnarray}
Here we use the following notations
\begin{equation}
p = \frac{(\gamma^{\rm b})^2-(\gamma^{\rm b}_{\rm in})^2}{(\gamma^{\rm b}_{\rm in})^2},
~\Lambda = \frac{4 \lambda \sigma}{\gamma^{\rm b}_{\rm in}},~\Delta 
= \frac{\gamma^2-\gamma_{\rm in}^2}{\gamma^2}.
\end{equation}

\begin{figure}
\includegraphics[scale=0.58]{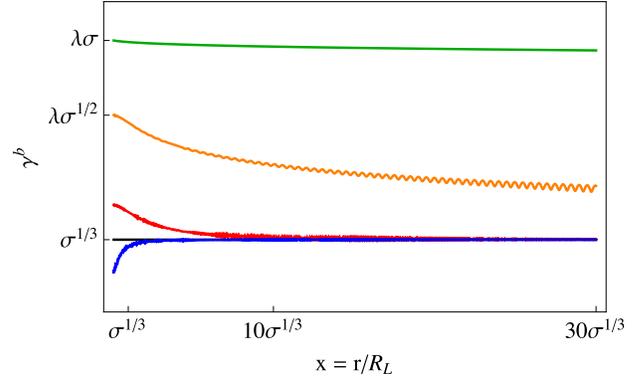}
\caption{
The evolution of the beam Lorentz-factor $\gamma^{\rm b}(x)$ for different initial energies
$m_{\rm e}c^2 \gamma_{\rm in}^{\rm b}$ (see text for more detail).
}
\label{fig:damp}
\end{figure}

 
The solutions of equation (\ref{eqn:damp}) for different initial energy 
$m_{\rm e}c^2 \gamma_{\rm in}^{\rm b}$ of a beam (and for 
$\gamma_{\rm in} = \gamma = \sigma^{1/3}$)  
are presented on Figure \ref{fig:damp}. As we see, for very large initial
Lorentz-factor of a beam $\gamma_{\rm in}^{\rm b} > \lambda \sigma$ 
(green line), 
its energy remains actually constant. The point is that the Larmor radius 
$r_{\rm L}  = m_{\rm e}c^2 \gamma^{\rm b}/eB_{\rm L}$ for such a large energy 
of a beam is much larger than the light cylinder radius $R_{\rm L}$ so the 
electromagnetic fields cannot disturb the radial motion of the beam. But, as 
we see, for $\gamma_{\rm in}^{\rm b} \approx \lambda \sigma^{1/2}$, when 
$r_{\rm L} \approx R_{\rm L}$ (orange line), the curve has oscillations 
corresponding to rotation in the comoving reference frame and beam energy 
slightly damps.

On the other hand, for smaller Lorenz-factors of the beam
\mbox{$\gamma_{\rm in } < \gamma_{\rm in}^{\rm b} < \lambda \sigma^{1/2}$} 
(red line) the beam energy marginally damps to the energy of secondary plasma. 
Here again the oscillations correspond to the rotation in the comoving reference 
frame. For $\gamma_{\rm in}^{\rm b} = \gamma = \sigma^{1/3}$ (black line), the 
energy of the beam remains unchanged. Finally, if the initial energy of the 
beam is relatively close to the energy of the secondary plasma, the beam either 
damps or accelerates to the secondary plasma energy. Although the case of 
$\gamma_{\rm in}^{\rm b}<\gamma$ (blue line) is impossible in real pulsars, 
this solution describes the behaviour of secondary particles: if their initial 
Lorentz-factor is less than $\sigma^{1/3}_{\rm M}$, they become accelerated to 
this energy. 

For sufficiently large initial Lorenz-factors of the beam, one can put $p \ll 1$ 
and neglect the second term in (\ref{eqn:damp}). In this case Eqn. (\ref{eqn:damp}) 
becomes linear, and one could solve it using Green's function:
\begin{equation}
\label{eqn:dump-sol}
p(x) = \Lambda^2\int\limits_{1}^{x}\text{d}x' \sin\left(\dfrac{\Lambda}{x}
-\dfrac{\Lambda}{x'}\right)\left[\dfrac{1}{\gamma^2(x')}+\dfrac{\Delta(x')}{x'^2}\right].
\end{equation}
After transition to the limit $x\rightarrow \infty$, we obtain the following 
asymptotic solution:
\be
\label{eqn:dump result}
p(x)\simeq-\left(\dfrac{\Lambda}{\gamma_{\rm out}}\right)^2[\ln(x/\Lambda)-\Gamma].
\ee
Here 
\be
\gamma_{\rm out}=\lim\limits_{x\rightarrow \infty}\gamma(x),
\ee
and $\Gamma\simeq 1$--$10$ is a constant which weakly depends on behavior of 
$\gamma(x)$ and $\Lambda$. For constant $\gamma$ and $\Lambda \ga 1$, this 
value is equal to the Euler-Mascheroni constant $\Gamma \approx 0.577$.

\begin{figure}
\includegraphics[scale=0.8]{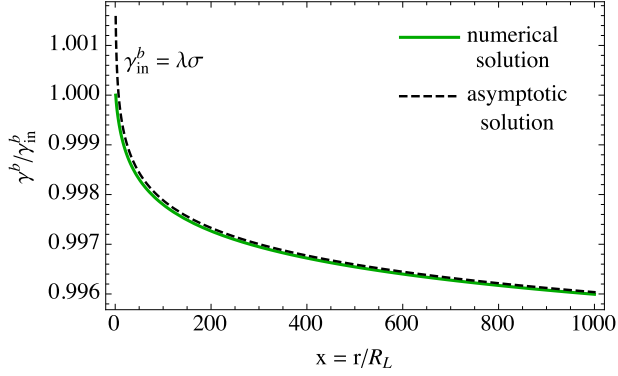}
\caption{Evolution of the Lorenz-factor of the beam for  
\mbox{$\gamma_{\rm in}^{\rm b} = \lambda \sigma$.} Green solid line 
represents numerical solution of exact equation (\ref{eqn:damp}) 
and black dashed line describes analytical asymptotic 
solution (\ref{eqn:dump result}).
}
\label{fig:sol}
\end{figure}

Approximating now $\ln(x/\Lambda)-\Gamma\sim1$, one can finally obtain the following 
equation describing Lorentz-factor of the primary particles:
\be
(\gamma^{\rm b}_{\rm out})^2 = (\gamma^{\rm b}_{\rm in})^2 - 
(4\lambda \sigma/\gamma_{\rm out})^2.
\ee
Hence, as was already found, large enough disturbance of the energy of the
primary beam $\Delta\gamma^{\rm b}\simeq\gamma^{\rm b}_{\rm in}$
is valid only if the condition $\gamma^{\rm b}_{\rm in} < \lambda\sigma^{1/2}$ 
holds. For  $\gamma^{\rm b}_{\rm in} > \lambda\sigma$ the energy of the
beam remains actually unchenged. Figure {\ref{fig:sol}} illustrates 
the evolution of $\gamma^{\rm b}(r)$ according to the numerical 
solution of Eqn. (\ref{eqn:damp}) in comparison with analytical solution 
(\ref{eqn:dump result}). As we see, they are in a very good agreement.
To summarize, the beam energy stabilizes at the following values:
\begin{eqnarray}
\gamma_{\rm out}^{\rm b} & = & \gamma_{\rm out};
~~~~~~~~\gamma_{\rm in}^{\rm b} < \lambda \sigma^{1/2}, \\
\gamma_{\rm out} < \gamma_{\rm out}^{\rm b} & < & \gamma_{\rm in}^{\rm b};
~~~~~~~~ \lambda \sigma^{1/2} < \gamma_{\rm in}^{\rm b} <  \lambda \sigma, \\
\gamma_{\rm out}^{\rm b} & = & \gamma_{\rm in}^{\rm b};
~~~~~~~~ \gamma_{\rm in}^{\rm b} >  \lambda \sigma,
\end{eqnarray} 
where $\gamma_{\rm out}$ 
is the final Lorentz-factor of the secondary plasma (see below).

Note, that if $\gamma_{\rm in}^{\rm b} > \lambda \sigma$, the pulsar total energy losses are less than the beam energy. This case is thus unphysical.

\subsection{Secondary plasma acceleration}

We are now in position to evaluate the final Lorentz-factor of the secondary 
plasma $\gamma_{\rm out}$. For this we look for the solution of equation (\ref{GE}) 
in the limit $x\rightarrow \infty$. Neglecting the disturbance of electric current 
we can approximate $K(r,\theta)$ in Eqn. (\ref{KD}) as
\begin{equation}
K_{\infty} = \delta_{\infty} -\frac{\gamma^{\rm b}_{\rm out}
-\gamma_{\rm in}^{\rm b}}{2\sigma\lambda}.
\end{equation}
At large distances outside the fast magnetosonic surface we can neglect 
two last terms in equation (\ref{GE}), which gives us a simple expression 
for $\gamma_{\rm out}$:
\be
\label{eqn:secondary}
\gamma_{\rm out} = \gamma_{\rm in}+ \sigma \left(\delta_\infty 
-\frac{\gamma^{\rm b}_{\rm out}-\gamma_{\rm in}^{\rm b}}{2\lambda\sigma } \right).
\ee
According to Eqns. (\ref{DelS}), (\ref{DelG}), 
$\delta_\infty \sim \max(\sigma^{-2/3}, \gamma_{\rm in} \sigma^{-1})$, so
we approximate equation (\ref{eqn:secondary}) as
\be
\label{eqn:gOut}
\gamma_{\rm out} = \max(\sigma^{1/3},\gamma_{\rm in})-\frac{\gamma^{\rm b}_{\rm out}
-\gamma_{\rm in}^{\rm b}}{2\lambda}.
\ee
In what follows we consider for simplicitly the case \mbox{$\gamma_{\rm in} \ll \sigma^{1/3}$,} 
but our results could be easily generalized for the opposite case. 

Note, that equation (\ref{eqn:gOut}) sets $\gamma_{\rm out}$ implicitly since 
$\gamma_{\rm out}^{\rm b}$ depends on $\gamma_{\rm out}$:
\be
\gamma^{\rm b}_{\rm out}-\gamma_{\rm in}^{\rm b} \approx 
\frac{(\gamma^{\rm b}_{\rm out})^2-(\gamma_{\rm in}^{\rm b})^2}{2 \gamma_{\rm in}^{\rm b}} 
= \frac{\gamma_{\rm in}^{\rm b}}{2}p_{\rm out}(\gamma_{\rm out}) .
\ee
This approximation is valid only when the beam damping is small, i.e., for large enough
$\gamma_{\rm in}^{\rm b}$.  In this limit, using equation (\ref{eqn:dump result}), we finally
obtain
\be
\label{eqn:gOutLarge}
\gamma_{\rm out} = \sigma^{1/3} + \frac{4 \lambda 
\sigma^2}{\gamma_{\rm in}^{\rm b} \gamma_{\rm out}^2}.
\ee
For very large $\gamma_{\rm in}^{\rm b}$, one can put 
$\gamma_{\rm out} = \gamma_{\rm in} = \sigma^{1/3}$, so both the beam and 
the secondary plasma do not change their energy. But there is a range of 
the value $\gamma_{\rm in}^{\rm b}$, where the second term in Eqn. 
(\ref{eqn:gOutLarge}) is larger than the first one. 
For such parameters
\be
\label{eqn:gOutLarge2}
\gamma_{\rm out} \sim \frac{\lambda^{1/3}
\sigma^{2/3}}{(\gamma_{\rm in}^{\rm b})^{1/3}}.
\ee
Finally, for sufficiently small $\gamma_{\rm in}^{\rm b}$, one can put 
\mbox{$\gamma_{\rm out}^{\rm b} =  \sigma^{1/3}$.} Then, for $\lambda \gg 1$          
equation (\ref{eqn:gOut}) transforms into
\be
\label{eqn:gOutSmall}
\gamma_{\rm out} = \sigma^{1/3} + \frac{\gamma_{\rm in}^{\rm b}}{2 \lambda}.
\ee
Thus, for small enough initial energy of the beam, the energy of the secondary particles 
linearly increases with $\gamma_{\rm in}^{\rm b}$
{\footnote{For $\gamma_{\rm in}^{\rm b} < \gamma_{\rm in}$,
equation (\ref{eqn:gOut}) gives $\gamma_{\rm out}<\gamma_{\rm in}$, i.e., the secondary 
plasma loses energy and accelerates the primary particles}}.

\begin{figure}
\includegraphics[scale=0.58]{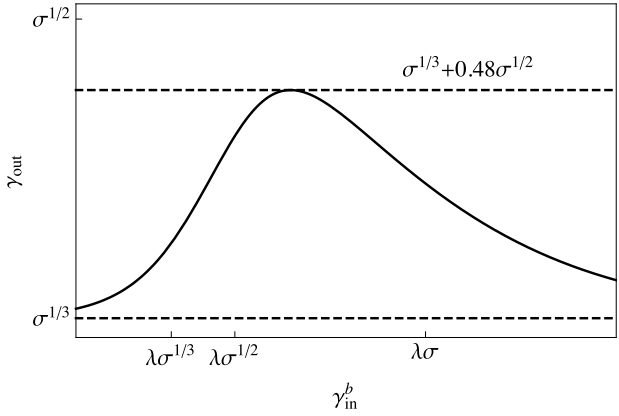}
\caption{The final Lorentz-factor of the secondary plasma according to Eqn. 
(\ref{eqn:final}). }
\label{fig:gOut}
\end{figure}

One can now construct the following approximate expression, which is valid in both limits
\be
\label{eqn:final}
\gamma_{\rm out} = \sigma^{1/3} + 
\left[\left(\frac{\gamma_{\rm in}^{\rm b}}{2 \lambda}\right)^{-1}
+ \left(\frac{\lambda^{1/3} \sigma^{2/3}}{(\gamma_{\rm in}^{\rm b})^{1/3}}\right)^{-1}\right]^{-1}
\ee
and determine the maximum gamma factor of the secondary plasma:
\be 
\gamma_{\rm out}^{\rm max} \sim \sigma^{1/2},~{\rm for}~\gamma_{\rm in}^{\rm b} 
= \lambda \sigma^{1/2}_{\rm M}.
\ee
Figure \ref{fig:gOut} shows obtained asymptotic behavior (\ref{eqn:final}). The exact 
location of the maximum is $\gamma_{\rm in}^{\rm b} = 6^{3/4} \lambda \sigma^{1/2}_{\rm M}$ 
and $\gamma_{\rm out}^{\rm max} = \sigma^{1/3}_{\rm M} + 0.25 (3/8)^{3/4} 
\sigma^{1/2}_{\rm M} \approx\sigma^{1/3}_{\rm M} +0.48\sigma^{1/2}_{\rm M} $.
One should also note that the maximum gamma-factor of the secondary plasma is close to the initial 
one: even for the fastest pulsars, 
\be 
\gamma_{\rm out}^{\rm max}/\gamma_{\rm in} = \sigma^{1/6} \sim 10 .
\ee
 
All the expressions written above remains the same in case $\gamma_{\rm in}\gg\sigma^{1/3}$
as well. The only differences is in first terms of r.h.s of Eqns. (\ref{eqn:gOutLarge}), 
(\ref{eqn:gOutSmall}), and (\ref{eqn:final}), which  are replaced on $\gamma_{\rm in}$ 
instead of $\sigma^{1/3}_{\rm M}$. This implies that for such pulsars the effect of 
the pulsar wind acceleration via beam particles is even smaller. For 
$\gamma_{\rm in}\gg \sigma^{1/2}_{\rm M}$ this effect vanishes.

\section{Conclusion}
\label{sect:res}

It was demonstrated that the effective deceleration of the primary beam takes place
only for relatively small beam energy
\begin{equation}
\gamma_{\rm in}^{\rm b} < \lambda \sigma^{1/2}.
\end{equation}
As $\lambda$ and $\sigma$ are connected with the relation \citep{2010mfca.book.....B}
\begin{equation}
\sigma \lambda \sim \left(\frac{W_{\rm tot}}{W_{\rm A}}\right)^{1/2},
\label{newsigma}
\end{equation}
where $W_{\rm tot}$ is the total energy losses of a pulsar and
\mbox{$W_{\rm A} = m_{\rm e}^{2}c^{5}/e^{2} \approx 10^{17}$ erg/s,}
one can easily estimate the total number of pulsars having effective damping 
of the primary beam. Setting $\gamma_{\rm in}^{\rm b} = 10^7$ and $\lambda = 10^4$, we 
obtain that among 2003 known radio pulsars with measured $P$ and  $\dot P$, only
65 sources have relatively small initial Lorentz-factors 
$\gamma^{\rm b}_{\rm in} < \lambda \sigma^{1/2}$.

Thus, one can conclude that effective deceleration of the primary beam takes place only 
for sufficiently energetic pulsars satisfying the condition $W_{\rm tot} \ga 10^{36}$ 
erg/s (Crab, Vela). In this case during its motion through the magnetosphere the primary 
beam decelerates, accelerating simultaneously the secondary particles. But such an 
additional acceleration is not actually effective enough as the maximum Lorentz-factor 
of the secondary plasma does not exceed $\sigma^{1/2}$, which is only on the factor 
$\sigma^{1/6}$ larger than the energy of the seconary plasma near the fast magnetosonic
surface. For ordinary pulsars the deceleration of the primary beam can be neglected, 
so the energy of the primary beam as well as secondary plasma remain constant.  

It is very important that the deceleration/acceleration takes place on the scale of the fast 
magnetosonic surface $r_{\rm F} \sim \sigma^{1/3} R_{\rm L}$, i.e., at the distances much
larger than the radius of the light cylinder. This implies that even for the fast young radio 
pulsars the energetic particles of the beam will intersect the magnetospheric current sheet 
as their velocities exceed the velocity of the secondary plasma which just determines
the radial velocity of the current sheet. Indeed, for fast pulsars the scale 
$r_{\rm F} \sim 10^{2} R_{\rm L}$ corresponds to several wave length of the current sheet.   

\section{Acknowledgments}

We thank A.Spitkovsky, A.Philippov and Y.N.~Istomin for their interest and useful discussions. 
This work was partially supported by Russian Foundation for Basic Research (Grant no. 
14-02-00831). 

{\small
\bibliographystyle{mn2e}
\bibliography{beam}

\begin{thebibliography}{}

\bibitem[\protect\citeauthoryear{{Arons}}{{Arons}}{1981}]{1981ApJ...248.1099A}
{Arons} J.,  1981, \apj, 248, 1099

\bibitem[\protect\citeauthoryear{{Beskin}}{{Beskin}}{2010}]{2010mfca.book.....B}
{Beskin} V.~S.,  2010, {MHD Flows in Compact Astrophysical Objects}.
Springer

\bibitem[\protect\citeauthoryear{{Beskin}, {Kuznetsova} \& {Rafikov}}{{Beskin}
  et~al.}{1998}]{bes98}
{Beskin} V.~S.,  {Kuznetsova} I.~V.,    {Rafikov} R.~R.,  1998, \mnras, 299,
  341

\bibitem[\protect\citeauthoryear{{Beskin} \& {Rafikov}}{{Beskin} \&
  {Rafikov}}{2000}]{2000MNRAS.313..433B}
{Beskin} V.~S.,  {Rafikov} R.~R.,  2000, \mnras, 313, 433

\bibitem[\protect\citeauthoryear{{Beskin}, {Zakamska} \& {Sol}}{{Beskin}
  et~al.}{2004}]{beszak04}
{Beskin} V.~S.,  {Zakamska} N.~L.,    {Sol} H.,  2004, \mnras, 347, 587

\bibitem[\protect\citeauthoryear{{Bogovalov}}{{Bogovalov}}{2001}]{2001A&A...371.1155B}
{Bogovalov} S.~V.,  2001, \aap, 371, 1155

\bibitem[\protect\citeauthoryear{{Contopoulos}, {Kazanas} \&
  {Fendt}}{{Contopoulos} et~al.}{1999}]{ckf99}
{Contopoulos} I.,  {Kazanas} D.,    {Fendt} C.,  1999, \apj, 511, 351

\bibitem[\protect\citeauthoryear{{Daugherty} \& {Harding}}{{Daugherty} \&
  {Harding}}{1982}]{daugherty1982pulsarcascades}
{Daugherty} J.~K.,  {Harding} A.~K.,  1982, \apj, 252, 337

\bibitem[\protect\citeauthoryear{{Gurevich} \& {Istomin}}{{Gurevich} \&
  {Istomin}}{1985}]{1985ZhETF..89....3G}
{Gurevich} A.~V.,  {Istomin} I.~N.,  1985, Zhurnal Eksperimentalnoi i
  Teoreticheskoi Fiziki, 89, 3

\bibitem[\protect\citeauthoryear{{Istomin} \& {Sobyanin}}{{Istomin} \&
  {Sobyanin}}{2007}]{2007AstL...33..660I}
{Istomin} Y.~N.,  {Sobyanin} D.~N.,  2007, Astronomy Letters, 33, 660

\bibitem[\protect\citeauthoryear{{Manchester} \& {Taylor}}{{Manchester} \&
  {Taylor}}{1977}]{1977puls.book.....M}
{Manchester} R.~N.,  {Taylor} J.~H.,  1977, {Pulsars}.
W.~H.~Freeman

\bibitem[\protect\citeauthoryear{{Medin} \& {Lai}}{{Medin} \&
  {Lai}}{2007}]{2007MNRAS.382.1833M}
{Medin} Z.,  {Lai} D.,  2007, \mnras, 382, 1833

\bibitem[\protect\citeauthoryear{Michel}{Michel}{1994}]{michel_pulsar_wind_1994}
Michel F.~C.,  1994, ApJ, 431, 397

\bibitem[\protect\citeauthoryear{{Philippov} \& {Spitkovsky}}{{Philippov} \&
  {Spitkovsky}}{2014}]{2014ApJ...785L..33P}
{Philippov} A.~A.,  {Spitkovsky} A.,  2014, \apjl, 785, L33

\bibitem[\protect\citeauthoryear{{Ruderman} \& {Sutherland}}{{Ruderman} \&
  {Sutherland}}{1975}]{rs75}
{Ruderman} M.~A.,  {Sutherland} P.~G.,  1975, \apj, 196, 51

\bibitem[\protect\citeauthoryear{{Smith}}{{Smith}}{1977}]{1977puls.book.....S}
{Smith} F.~G.,  1977, {Pulsars}.
Cambridge University Press

\bibitem[\protect\citeauthoryear{{Spitkovsky}}{{Spitkovsky}}{2006}]{2006ApJ...648L..51S}
{Spitkovsky} A.,  2006, \apjl, 648, L51

\bibitem[\protect\citeauthoryear{{Sturrock}}{{Sturrock}}{1971}]{sturrock71}
{Sturrock} P.~A.,  1971, \apj, 164, 529

\bibitem[\protect\citeauthoryear{{Tchekhovskoy}, {Spitkovsky} \&
  {Li}}{{Tchekhovskoy} et~al.}{2013}]{SashaMHD}
{Tchekhovskoy} A.,  {Spitkovsky} A.,    {Li} J.~G.,  2013, \mnras, 435, L1

\bibitem[\protect\citeauthoryear{{Timokhin} \& {Arons}}{{Timokhin} \&
  {Arons}}{2013}]{2013MNRAS.429...20T}
{Timokhin} A.~N.,  {Arons} J.,  2013, \mnras, 429, 20

\bibitem[\protect\citeauthoryear{{Tomimatsu}}{{Tomimatsu}}{1994}]{1994PASJ...46..123T}
{Tomimatsu} A.,  1994, \pasj, 46, 123

\bibitem[\protect\citeauthoryear{{Zheleznyakov}}{{Zheleznyakov}}{1977}]{1977ewcp.book.....Z}
{Zheleznyakov} V.~V.,  1977, {Electromagnetic waves in cosmic plasma.
  Generation and propagation}.
Nauka

\end{thebibliography}
}
\bsp

\appendix
\section{Radiation force damping}
\label{LC}

Inside the light cylinder the damping of a beam is defined by radiation losses. 
Outside the acceleration gap the change of the bulk Lorentz-factor of a beam is 
described by following equation~\citep{1977ewcp.book.....Z}
\be
\label{RL}
mc^2\frac{{\rm d}\gamma}{{\rm d}l}=-\frac{2}{3}\frac{e^2}{R_c^2}\gamma^4,
\ee
where $R_c$ is the curvature radius of particle trajectory (i.e., the curvature
radius of the magnetic field line), and $l$ is a path. For simplify we consider 
here a dipole structure of magnetic field inside the light cylinder. 
 
Dipole magnetic field lines in polar coordinates are described by the well-known relation
$r=r_0\sin^2\theta$, where \mbox{$r_0 \ga R_{\rm L}$} is a parameter of the line. The appropriate 
expression in Cartesian coordinates looks like $y=\pm\sqrt{(r_0 x^2)^{2/3}-x^2}$. It gives 
for the curvature radius
\be
\label{CR}
R_c=r_0\dfrac{a^{1/3}(4-3a^{2/3})^{3/2}}{(6-3a^{2/3})},
\ee 
where $a=x/r_0$. Accordingly, 
\be
\label{al}
{\rm d}l=\dfrac{r_0}{3}\sqrt{\dfrac{4-3a^{2/3}}{a^{2/3}(1-a^{2/3})}}{\rm d}a.
\ee
It gives
\be
\frac{\text{d}\gamma}{\gamma^4}=
-\frac{6r_e}{r_0}\int\limits_{R/r_0}^{R_L/r_0}\dfrac{{\rm d}a}{a}
\dfrac{(2-a^{2/3})^2}{(4-3a^{2/3})^{5/2}(1-a^{2/3})^{1/2}}.
\ee
Here $r_{\rm e} = e^2/m_{\rm e}c^2$ is the classical electron radius, and $R$ is the stellar 
radius. This equation can be easily solved analytically, and we obtain
\be
\frac{1}{\gamma_{0}^3}-\frac{1}{\gamma_{\rm in}^3}
=24 \, \dfrac{r_{\rm e}}{R_L}\left[2\left(\ln\frac{R_L}{R}+1\right)+F(r_0,R)\right],
\label{ggg}
\ee
where $F(r_0,R) \approx 1$ and, hence, may be neglected. 
As the initial Lorentz-factor of the primary beam after acceleration inside 
thin gap is estimated as $10^8$--$10^{9}$~\citep{2010mfca.book.....B}, one 
can neglect the second term in the l.h.s. of Eqn. (\ref{ggg}), and we finlly obtain
\be
\gamma_0 \sim 0.1 \left(\frac{R_{\rm L}}{r_{\rm e}}\right)^{1/3} \sim 10^7. 
\ee 

\section{Beam instability}
\label{BI}

In this section we consider the effects leading to the beam instability 
and estimate the distance were the beam deceleration by the wave-particle 
interaction takes place. For simplicity, we use the following expression for 
plasma dielectric permittivity
\begin{equation}
\varepsilon(\omega,{\bf k})=
1-\frac{(\omega_{\rm p})^2}{\gamma^3(\omega-{\bf k} {\bf v}_{\rm p})^2}
-\frac{(\omega_{\rm b})^2}{(\gamma^{\rm b})^3(\omega-{\bf k} {\bf v}_{\rm b})^2},
\end{equation}
where ${\bf v}_{\rm p}$ and ${\bf v}_{\rm b}$ are the velocities of the secondary plasma 
and the primary beam respectively, and $\omega_{\rm p}$ and $\omega_{\rm b}$ are 
corresponding plasma frequencies:
\begin{equation}
\omega_{\rm p,b}^2 = \frac{4 \pi e^2 n_{\rm p,b}}{m_e}.
\end{equation}

For ${\bf k} \parallel {\bf v_b} \parallel {\bf v_p}$, the spectrum of plasma 
oscillations (i.e., Langmuir waves) corresponds to the condition
$\varepsilon(\omega, {\bf k})=0$.  Considering now the solution as 
\begin{equation}
\omega = {\bf k} {\bf v}_{\rm b} + \delta\omega,
\end{equation}
where $\delta\omega \ll  {\bf k}{\bf v}_{\rm b}$, one can obtain for the maximum growth 
increment
\begin{equation}
\delta\omega_{\rm max} = 
\kappa\frac{(\omega_{\rm p})^{1/3}(\omega_{\rm b})^{2/3}}{\gamma^{1/2} \gamma^{\rm b}}
\end{equation}
with $\kappa \approx 1$.

Before becoming unstable, the beam travels the distance
\begin{equation}
d_{\rm stab}= \frac{c}{\delta\omega_{\rm max}} \approx 
\frac{c}{\omega_{\rm GJ}} \frac{\gamma^{1/2}\gamma^{\rm b}}{\lambda^{1/6}},
\end{equation}
where $\omega_{\rm GJ}^2 = 4\pi n_{\rm GJ} e^2/m_e = 2 \Omega \omega_{B}$. It gives
\begin{align}
\frac{d_{\rm stab}}{R_{\rm L}} &= \left(\frac{\Omega}{\omega_B}\right)^{1/2} 
\frac{\gamma^{1/2}\gamma^{\rm b}}{\lambda^{1/6}} = \\
&= 1500 B_{6}^{-1/2} P_{\rm s}^{-1/2}\gamma^{\rm b}_{7} \gamma_{2}^{-1/2} 
\lambda_{6}^{-1/6} \gg 1. \nonumber
\end{align}
Here we assume the value of the magnetic filed near the light cylinder to be of order 
$10^6$~G which is natural for fast radio pulsars. Thus, one can conclude that one can 
neglect the two-fluid instability within distances which were considered in this paper.

One can find even more precise expressions considering the values 
$\omega_{\rm b},~\omega_{\rm p}\propto r^{-1}$ as depending on the
radius $r$, as it takes place in Michel solution 
(\ref{leqf1})-(\ref{leqf2}). The destruction distance could then be found from
\be
\int\limits_{R_{\rm L}}^{R_{\rm L} 
+ d_{\rm stab}}\frac{\omega_{\rm p}^{2/3}(r)
\omega_{\rm b}^{1/3}(r)}{\gamma( \gamma^{\rm b})^{1/2}}\frac{\text{d}r}{c} = 1.
\ee
Assuming now that $\gamma_{\rm p,b}$ do not depend on $r$, we obtain
\begin{align}
\log&\left(\frac{R_{\rm L} + d_{\rm stab}}{R_{\rm L}}\right) 
= \frac{c \gamma\gamma^{\rm b}}{R_{\rm L}\omega_{\rm p}^{2/3}(R_{\rm L})\omega_{\rm b}^{1/3}(R_{\rm L})} = \\
& = 1500 B_{8}^{-1/2} P_{\rm s}^{-1/2}\gamma^{\rm b}_{7} \gamma_{2}^{-1/2} 
\lambda_{6}^{-1/6} \gg 1. \nonumber
\end{align}
Thus, in reality not only the condition $d_{\rm stab}/R_{\rm L} \gg 1$, but much more 
strong one
\be
\log(d_{\rm stab}/R_{\rm L}) \gg 1
\ee
is valid.

\bsp

\label{lastpage}

\end{document}